\documentclass[12pt]{iopart}

\usepackage{iopams}

\usepackage{graphicx}

\newcommand{\AmS}{{\protect\the\textfont2
  A\kern-.1667em\lower.5ex\hbox{M}\kern-.125emS}}

  \newcommand \beq{\begin{eqnarray}}
\newcommand \eeq{\end{eqnarray}}
\newcommand{\bea}{\begin{eqnarray}}
\newcommand{\eea}{\end{eqnarray}}

\def\simle{\mathrel{\rlap{\raise 0.511ex \hbox{$<$}}{\lower 0.511ex \hbox{$\sim$}}}}
\def\simge{\mathrel{ \rlap{\raise 0.511ex \hbox{$>$}}{\lower 0.511ex \hbox{$\sim$}}}}

\def\0{\over } \def\2{{1\over2}} \def\4{{1\over4}}
\def\5{\hat } \def\6{\partial }

\def\8#1{{\textstyle{#1}}}

\def\({\left(} \def\){\right)} \def\<{\langle } \def\>{\rangle }

\begin{document}

\title{Theoretical overview: towards understanding the quark-gluon plasma}

\author{Jean-Paul Blaizot}

 \address{ECT*, Villa Tambosi, \\
        strada delle Tabarelle, 286, I 38050 Villazzano (TN), Italy}%

\ead{blaizot@ect.it}

\begin{abstract}
I give a brief overview of recent theoretical developments
concerning the high temperature phase of QCD, and the structure of
hadronic wave functions at high energy.
\end{abstract}

\section{From the ``ideal gas'' to the ``perfect liquid''}

The study of ultra-relativistic heavy ion collisions offers the possibility to address several fundamental questions, concerning for instance the state  of matter at very high temperature and density, or the structure of the wave-function of a nucleus at asymptotically high energy.
The reason why these questions refer to extreme situations is  of course the fact that  simplicity often emerges  in asymptotic situations, allowing for a deeper theoretical understanding, that can be eventually extrapolated to the more complex, non asymptotic, situations.

The naive picture of the quark-gluon plasma belongs to such
asymptotic idealizations: as a  natural consequence of  the QCD
asymptotic freedom, one expects hadronic matter  to turn at high
temperature and density into a gas of quarks and gluons whose free
motion is only weakly perturbed by their interactions.  The
beautiful RHIC data that have been collected over the last few years
have somehow shaken  our hope that such an idealized state of matter
can be observed in nuclear collisions: the temperature reached is
presumably not  high enough, or is attained for too short a period
of time to lead to observable consequences. Still,  a consensus has
been reached  that some form of a quark-gluon plasma is produced in
RHIC collisions (see
\cite{Arsene:2004fa,Back:2004je,Adcox:2004mh,Adams:2005dq}, and also
\cite{Gyulassy:2004zy,Shuryak:2004cy}). This has created a shift of
paradigm: we are no longer focusing on the  discovery of the
quark-gluon plasma, but rather on studying  its properties, that
RHIC allows us to   explore experimentally. Hence the title of this
talk,  where I have to emphasize, however, that our ``understanding"
is hampered by the fact that RHIC forces us to look in  a regime
where, unfortunately, theory is  hard.

The impact of the RHIC discoveries on current theoretical
investigations has been in fact even deeper, with most of the
present discussions emphasizing the strongly coupled character of
the quark-gluon plasma.  Three elements have actually conspired to
this further shift in paradigm. First, as  we have alluded to
already, the RHIC data do not provide any evidence for ideal gas
behavior. They are better interpreted by assuming that the produced
matter behaves as a liquid with low viscosity, the ``perfect liquid"
\cite{Shuryak:2003xe,Teaney:2003kp}. Second, perturbation theory is
notoriously unable to describe the quark-gluon plasma unless the
temperature is extremely high. Third, new techniques have emerged
that allow calculations to be done in some strongly coupled gauge
theories (that differ however in essential aspects from QCD). The
strongly coupled plasma becomes then  another idealized picture, of
the kind referred to earlier, that could be used as a starting point
for the description of the physical quark-gluon plasma.

When discussing the quark-gluon plasma in the context of heavy ion
collisions, one has to face the question of how such a state of
matter can be produced. This requires understanding the structure of
the wave function of a nucleus at very high energy, and the detailed
microscopic mechanisms by which its partonic degrees of freedom get
liberated and subsequently interact to lead possibly  to a
thermalized system. These issues will be very briefly discussed in
the last part of this talk. It is interesting to observe in this
context the remarkable merging of scientific interests in the small
$x$ physics
 and the physics of ultra-relativistic heavy ions, two fields
that until recently had little in common: in both domains one is
looking into regimes of QCD where the parton densities are large and
the non linearities of QCD play a major role.

\section{Is the quark-gluon plasma weakly or strongly coupled?}

This question does not have a straightforward answer. Indeed, as we
shall see, in the quark-gluon plasma coexist seemingly perturbative
features, and non perturbative ones. For instance lattice
calculations, which provide the most reliable information on the
properties of the quark-gluon plasma at high temperature, show that,
at very high temperature, the thermodynamical functions go, albeit
slowly, towards those of free massless particles (see for instance
\cite{Karsch:2001cy}), confirming the expected picture based on
asymptotic freedom. At the same time these calculations point to
important corrections to naive perturbation theory in all the
relevant range of temperatures.

\subsection{Breakdown of strict perturbation theory}

 Much effort has been put into calculating
 the successive orders of the perturbative expansion for the pressure
\cite{Arnold:1995eb,Zhai:1995ac,Braaten:1996jr,Kajantie:2002wa} and
the series is known now up to order $g^6\ln
g$\cite{Kajantie:2002wa}.  These calculations have revealed  that
perturbation theory  makes sense only for very small values of the
coupling constant, corresponding to extremely large values of $T$.
For not too small values of the coupling, the successive terms in
the expansion oscillate wildly and the dependence of the results on
the renormalization scale keeps increasing order after order (see e.g.
\cite{Blaizot:2003tw}). Clearly, the corrections to the ideal quark
gluon plasma cannot be calculated in strict perturbation theory.

 This situation is to be contrasted with what happens at zero temperature,
 where perturbative calculations achieve a reasonable accuracy already at the GeV scale.
 The point is that  the validity of a weak coupling expansion depends
not only on the strength of the coupling, but also on the number of
active degrees of freedom. At zero temperature, one deals most of
the time with a very limited number of degrees of freedom (the
colliding particles and the reaction products), while at finite
temperatures, as we shall see shortly, the  thermal fluctuations
alter the infrared behavior in a profound way. There are of course
situations at zero temperature where one needs to take into account
many degrees of freedom through appropriate resummations; the small
$x$ resummations provide an example. Similarly, in the study of cold
dense matter, the BCS instability towards color-superconductivity
provides another example of a weak coupling situation made non perturbative by the presence of many
degenerate degrees of freedom.
Having recognized this, we shall see that a lot can be learned from
weak coupling calculations, in particular how to capture the right
physics that can allow for smooth extrapolations towards the strong
coupling regime.

\subsection{The role of thermal fluctuations}

A simple characterization of the strength of the interactions in classical plasmas is provided by the  dimensionless parameter  $g^2\equiv e^2 n^{1/3}/T$, which is essentially the ratio  between   the average potential energy per particle ($\sim e^2 n^{1/3}$) and the mean kinetic energy  per particle ($\sim T$), with $T$  the temperature,   $n$  the number density and $e$ the electric charge. The condition that the plasma be ideal, or weakly interacting,  is then $T\gg e^2 n^{1/3}$, or simply $ g\ll 1$.
The Debye screening length is
$
\sim\sqrt{{T}/{ne^2}}\sim n^{-1/3}/g$  which,
 when $g$ is small,  is large compared
 to the interparticle distance ($\sim n^{-1/3}$), a criterion for collective behaviour.

In ultrarelativistic plasmas, the temperature and the density are no
longer independent control parameters since   $n\sim T^3$, and in
QCD,    the parameter $g$ reduces to  the gauge coupling (at a scale
$\sim T$). Note that, just above $T_c$, $g$ is not small, but not
huge either, $g\simeq 2$ (see e.g. \cite{Laine:2005ai}). However, as
already mentioned, to decide whether the quark-gluon plasma is
strongly or weakly coupled it  is  not enough to consider only the
strength of the coupling:  the effects of the interactions depend
also on the magnitude of the relevant thermal fluctuations, which
depends on their wavelengths.  Predicting the effect of the
interactions amounts to comparing the kinetic energy $\sim
\langle(\partial A)^2\rangle\sim k^2\langle A^2\rangle$ with the
interaction energy $g^2\langle A^4\rangle\sim g^2 \langle
A^2\rangle^2$, or equivalently $k^2$ with $g^2\langle A^2\rangle$
($A$ is the gauge field). At weak coupling a hierarchy of scales of
thermal fluctuations emerges, each scale being associated with well
identified physics (see e.g.  \cite{Blaizot:2001nr} for a more
detailed discussion). At one end we have the thermal fluctuations
corresponding to the   plasma particles  $\langle A^2\rangle_T\sim
T^2$. These constitute the  dominant contribution to the  energy
density at weak coupling. The plasma particles  have typical momenta
of the order of the temperature, and mostly kinetic energies,
$k^2\sim T^2\gg g^2T^2$ if $g$ is small.  For them, perturbation
theory works as well as at zero temperature. At the other end, we
have long wavelength ($\sim 1/g^2 T$), unscreened, magnetic
fluctuations which remain strongly coupled ($\langle
A^2\rangle_{g^2T}\sim g^2T^2$, so that $k^2\sim g^4 T^2\sim
g^2\langle A^2\rangle_{g^2T}$), however small the coupling, hence
however high the temperature. These magnetic fluctuations prevent
the perturbative calculation  of the pressure at order $g^6$ and
beyond.

Thus in non abelian ultra-relativistic plasmas, some degrees of
freedom can be weakly coupled while others remain strongly coupled
for any values of the coupling. The contribution of the magnetic
fluctuations to the pressure is not known quantitatively, although
there are indications from dimensional reduction (see below) that it
is small. Then the  main difficulty with thermal perturbation
theory, as far as the calculation of the pressure is concerned,
occurs already in scalar field theories \cite{Blaizot:2003tw}. It is
not so much related to the fact that the coupling is not small
enough (for the relevant temperatures the coupling is not huge), but
rather to the interplay of degrees of freedom with various
wavelengths, possibly involving collective modes. To deal with this
aspect of the problem, weak coupling techniques are useful: they
allow us to identify and perform  the appropriate reorganizations
and resummations of the perturbative expansion. Once this is done,
the extrapolation to strong coupling is much smoother than what
strict perturbation theory could lead us to expect.

\subsection{Effective theories and resummations}

 A powerful technique to handle situations where modes at different scales couple is that of effective theories. Among those, one focuses on the  modes carrying zero Matsubara frequencies. The construction of the effective theory for these modes  is known as dimensional reduction \cite{Appelquist:1981vg,Nadkarni:1983kb,Nadkarni:1988fh}. The coefficients of the resulting effective lagrangian are usually determined perturbatively as a function of the gauge coupling $g$ by matching. Calculations based on this scheme have been pushed to high order \cite{Kajantie:2000iz}, but the determination of the order $g^6$ contribution to the pressure  depends on an as yet undetermined 4-loop matching
coefficient. By adding a parameter to account for this uncalculated contribution, one can match the four-dimensional lattice results. The required value of the coefficient is not very large, suggesting that the contribution of the magnetic sector to the pressure is numerically not important at high temperature, as mentioned before.

Other ways to reorganize the perturbative expansion have been tried.
One proposal, called ``screened
perturbation theory'' \cite{Karsch:1997gj,Andersen:2000yj},  has
been generalized to the resummation of the full QCD hard thermal loops \cite{Andersen:1999fw,Andersen:2002ey}. A different approach, borrowed from early studies of another strongly interacting liquid, namely liquid helium, is based on  an expression for the
entropy density that can be obtained from a $\Phi$-derivable
two-loop approximation \cite{Blaizot:1999ip,Blaizot:2000fc} (sometimes also called 2PI formalism). This approach has also the virtue of providing a clear physical picture:  the dominant effect of the interactions is to turn  the original degrees of freedom, quarks and gluons, into massive quasiparticles, with weak residual interactions.  It was shown in
Refs.~\cite{Blaizot:1999ip,Blaizot:2000fc} that the lattice results for
the entropy of the gluonic  plasma  \cite{Boyd:1996bx,Okamoto:1999hi}  were quite well
reproduced  for $T \ge 3 T_c$. The  formalism used in this calculation of the entropy    has been tested \cite{Blaizot:2005wr} in the limit of a large number of quark flavors, which can be solved exactly. One then found that the 2PI approximation scheme allows for a smooth extrapolation that is accurate up to quite large values of the coupling constant.

The weak coupling approximation schemes can be fully justified only
when the coupling is truly smalI, in which cases the various
relevant degrees of freedom can be arranged in a clean hierarchy of
scales ($T, gT, g^2T$). The non perturbative renormalization group
\cite{Wetterich:1992yh,Ellwanger:1993kk,Tetradis:1993ts,Morris:1993qb,Morris:1994ie}
sheds some light on what happens as the coupling grows.  There is
some analogy between the effective field theory approach and the non
perturbative renormalization group: in effective field theory one
integrates out degrees of freedom above some cut-off; in the
renormalization group this integration is done smoothly.   In a way,
the renormalization group builds up a continuous tower of effective
theories that lie infinitesimally close to each other, and  are
related by a renormalization group flow equation.  This picture is
independent of the value of the coupling, so that the
renormalization group provides  a smooth extrapolation from the
regime of weak coupling, characterized by a clean separation of
scales, towards the strong coupling regime where all scales get
mixed.  In a recent study \cite{BIMW}, it has been shown in the case
of a scalar  $\phi^{4}$ theory   with $O(N)$ symmetry, that such a
technique can provide  a smooth extrapolation to strong coupling,
which turns out to be similar to  that of a simple 2PI
approximation.

The physical picture that emerges from these various weak coupling
calculations, with proper resummations, is that at high $T$, the
dominant degrees of freedom are quark and gluon quasiparticles. The
main effect of the interactions is to affect the propagation of
these quasiparticles that suffer otherwise very little residual
interactions. This picture is consistent with lattice calculations
for temperatures above $ 3T_c$. The physics  in the temperature
range $T_c\simle T\simle 3T_c$  remains poorly understood, with many
unanswered questions, for instance:  What are the degrees of freedom
relevant for an effective theory? Are remnants of confinement
important, such as those captured by  effective theories for the
Polyakov loop (see for instance
\cite{Pisarski:2000eq,Vuorinen:2006nz} or, on a more
phenomenological side, \cite{Ratti:2005jh})? What is the role, if
any, of  bound states \cite{Shuryak:2004tx}? What is their fate as
the temperature grows, do they survive at large temperature, as
recent lattice data suggest \cite{Asakawa:2003re}? What are the
charge carriers \cite{Ejiri:2005wq,Gavai:2005yk}? Etc.

\subsection{The strong coupling regime from the AdS/CFT correspondence}

New techniques for doing calculations in strongly coupled gauge theories have emerged recently. These techniques are based on a duality between   some supersymmetric Yang-Mills theories and
gravitation theories. The duality involves an interchange of the
regimes of weak and strong coupling: weak coupling in the
gravity theory corresponds to strong coupling in the gauge
theory. This duality offers the possibility to study strongly coupled gauge theories by performing essentially classical calculations in their gravity duals.  This possibility has been exploited in a number
of recent publications (see the talk by Hong Liu at this conference). One  prediction of such calculations concerns the entropy $S$ which behaves, in strong coupling, as \cite{Gubser:1998nz}
\beq
\frac{S}{S_0}=\frac{3}{4}+\frac{45}{32}\zeta(3)\frac{1}{\lambda^{3/2}}.
\eeq Here $\lambda\equiv 2g^2 N_c$ and $S_0$ is the entropy of the non interacting system. Thus,
 in the limit of strong coupling, $\lambda\to\infty$,
the entropy is bounded from below by the value $S/S_0=3/4$. The fact
that this value is close to that obtained in lattice calculations at
temperatures above $T_c$ has led to the suggestion that the
quark-gluon  plasma above $T_c$ could be in  a strongly coupled
regime analogous to the strong coupling regime of SYM theories.
Comparing the two theories is, however, delicate. Note first that ${\cal N}=4$ SYM theories, the most discussed of such theories, and the only ones that I consider in this short discussion, have symmetries that make them
rather different from QCD: in particular the coupling constant does
not run, and there is no phase transition. The running of the
coupling constant is however an essential property of QCD; in
particular, above the transition region  it is accompanied by a
breaking of conformal symmetry that is very well observed in lattice
calculations \cite{Boyd:1996bx,Gavai:2004se,Gavai:2005da}. So one
cannot compare ${\cal N}=4$ SYM theories with QCD in the temperature range where
the QCD coupling is the largest (the  minimal value of $3/4$ is
obviously not compatible with lattice data near $T_c$ where the
entropy nearly vanishes, and the coupling is the strongest). Thus
one should compare ${\cal N}=4$ SYM theories with QCD only for temperatures large
enough for conformal symmetry to be appproximately satisfied, that
is for $T\simge  3T_c$. But in this temperature range,  the entropy
density is about half-way between its weak coupling value and its
strong coupling value, so there is no compelling reason to favor at
that point an interpretation  based  on strong coupling (see
\cite{Blaizot:2006tk} for a recent discussion).

\section{Hadronic wave-functions at high energy and the early stages of nucleus-nucleus collisions}

I turn now to the second fundamental issue  mentioned in the
introduction, that related to the structure of the wave functions of
nuclei at very high energies. This is a domain where significant
progress has been achieved recently. I should emphasize that because
of lack of space and time, I shall not be able to do justice to
these exciting developments. I wish to refer to the talks by F.
Gelis and M. Strickland at this conference for more information. As
we shall see, much of the discussion relies on weak coupling
arguments, although strong interactions are generated by classical
color fields.

\subsection{The physics of saturation and the color glass condensate}

The degrees of freedom involved in the early stages of a
nucleus-nucleus collision at sufficiently high energy are partons,
mostly gluons, whose density grows as the energy increases (i.e.,
when $x$, their momentum fraction, decreases). This  phenomenon
 has been well established at HERA \cite{Chekanov:2002pv}. One
expects however that the growth of the gluon density should
eventually ``saturate'' when non linear QCD effects start to play a
role.

The existence of such a saturation regime has been predicted long
ago\cite{Gribov:1984tu,Mueller:1985wy}, together with estimates for
the typical transverse momenta where it could set in in heavy ion
collisions \cite{Blaizot:1987nc}. But it is only during the last
decade that equations providing a dynamical description of the
saturated regime have been obtained
\cite{Balitsky:1995ub,Kovchegov:1999ua} and
 \cite{Jalilian-Marian:1997jx,Kovchegov:1996ty,Iancu:2000hn,Weigert:2000gi,Ferreiro:2001qy}. A remarkable feature which emerges from the
solution of these equations is that the dense, saturated systems of
partons that make hadronic wave functions at high energy have
universal properties (see below), the same for all hadrons or
nuclei.

The momentum scale $Q_s$ that characterizes the onset of saturation
is called the saturation momentum  and is given by $Q_s^2 \sim
\alpha_s(Q_s^2){xG(x,Q_s^2)}/{\pi R^2}$ \cite{Mueller:1999wm}.
Partons in the wave function have different transverse momenta
$k_T$. Those with  $k_T> Q_s$ are in a dilute regime; those with
$k_T<Q_s$ are in the saturated regime. Note that at saturation,
naive perturbation theory breaks down, even though $\alpha_s(Q_s)$
may be small if $Q_s$ is large: the saturation regime is a regime of
weak coupling, but large density. In fact one can easily estimate
the number of partons occupying a small disk of radius $1/Q_s$ in
the transverse plane. At saturation, this number is proportional to
$ 1/\alpha_s$, a large number is $\alpha_s$ is small.  In such
conditions of large numbers of quanta, classical field
approximations may become relevant to describe the nuclear
wave-functions. This observation is at the basis of the
McLerran-Venugopalan model \cite{McLerran:1993ni}. The color glass
formalism provides a more complete physical picture. It relies on
the separation of the degrees of freedom into   Lorentz contracted
 frozen color sources $\rho$ flying along the light-cone, and low $x$
partons which are described by classical gauge fields $A^\mu(x)$
determined by solving the Yang-Mills equations with the source
$\rho$.  An average over all acceptable configurations must be
performed in order to calculate observables. In this average, the
weight of a given configuration is a functional $W_{x_0}[\rho]$ of
the density $\rho$ of color sources, which depends on the separation
scale $x_0$ between the modes which are treated as sources, and
those which are treated as fields. As one lowers this separation
scale, more and more modes are included among the frozen sources,
and therefore the functional $W_{x_0}$ evolves with $x_0$ according
to a renormalization group equation. This leads to the non  linear
evolution equations alluded to earlier, namely the
Balitsky-Kovchegov equation \cite{Balitsky:1995ub,Kovchegov:1999ua}
and the JIMLWK equation
 \cite{Jalilian-Marian:1997jx,Kovchegov:1996ty,Iancu:2000hn,Weigert:2000gi,Ferreiro:2001qy}.

The saturation momentum increases as the gluon density increases.
This may come from an increase of the gluon structure function as
$x$ decreases ($Q_s^2\sim x^{-0.3}$). This may also come from the
additive contributions of several nucleons in a nucleus. In large
nuclei, one expects $xG_A(x,Q_s^2) \propto A$, and hence
$Q_s^2\propto\alpha_s A^{1/3}$, where $A$ is the number of nucleons
in the nucleus. Thus, the saturation regime sets in earlier (i.e.,
at lower energy) in collisions involving large nuclei than in those
involving protons. In fact, the parton densities in the central
rapidity region of a Au-Au collision at RHIC are not too different
from those measured in deep inelastic scattering at HERA. The study
of $dA$ collisions at RHIC, in the fragmentation region of the
deuteron, gives access to a regime of smaller $x$ values, where
quantum evolution could be significant. Indeed, very exciting
results have been obtained in this regime \cite{Arsene04}, which
have been interpreted as evidence of saturation (see e.g.
\cite{Kharzeev:2004yx}).

\subsection{Fluctuations}

The non linear JIMWLK or BK evolution equations have solutions that
exhibits universal asymptotic properties (when the evolution is
carried to high rapidity) \cite{IIM02}. One   remarkable property is
that of geometrical scaling, observed at HERA \cite{GolecW12}. It
has been recognized recently that those asymptotic properties are
those of a large universality class of non linear equations that
appear in various areas of statistical physics, and describe
typically reaction-diffusion processes \cite{MP03}.

This connection with problems in statistical mechanics has been
deepened. In particular, the role of fluctuations in the dilute part
of the partonic systems has been recognized \cite{Iancu:2004es}.
Such fluctuations are ignored in the usual non linear evolution
equations, which emphasize the saturation regime where the large
densities validate a mean field approximation. New evolution
equations that incorporate the physics of fluctuations have been
constructed \cite{Iancu:2004iy,Mueller:2005ut}. A remarkable outcome
is that the fluctuations in the dilute regime influence the whole
approach to saturation.
 As a result, the
saturation scale becomes a fluctuating quantity, which, in a given
event, may reach different values for different impact parameters.
This could have phenomenological consequences at the very high
energies of the LHC (although perhaps no so much in A-A collisions,
since  the large size of the system will, a priori, damp then the
possible fluctuations).
 A preliminary study of such possible phenomenological consequences is given in \cite{Iancu:2006uc}.

\subsection{Initial conditions and the approach to thermalization}

The knowledge of the initial parton distribution is obviously
essential in order to determine the early stages of heavy ion
collisions. Typically  the partons which are set free during the
collision are those which carry a transverse momentum of order
$Q_s$,  and they are freed on a time scale of order $1/Q_s$. In the
``bottom-up'' scenario \cite{Baier:2000sb}, these hard gluons
radiate a lot of soft gluons and thermalization proceeds through
these soft gluonic modes that eventually collect all the energy
initially stored in the hard gluons. Such a scenario has been
revised recently \cite{Mueller:2005hj,Bodeker:2005nv} in order to
take into account the plasma instabilities generated by the
anisotropy of the initial momentum distributions \cite{Mrow94}.
These instabilities show up in many numerical simulations of the
initial stages of nucleus-nucleus collisions
\cite{Arnold:2004ti,Rebhan:2004ur,Romatschke:2006nk}. Whether they
play an essential role in the thermalization, or lead  perhaps to
turbulent behaviour
\cite{Arnold:2005qs,Mueller:2006up,Dumitru:2006pz}, or anomalous viscosity \cite{Asakawa:2007gf}, are issues, among
others, very much discussed at the moment.


\end{document}